\documentclass[conference,a4paper]{IEEEtran}

\usepackage{cmap}
\usepackage[utf8]{inputenc}
\usepackage[english]{babel}

\usepackage{indentfirst}
\usepackage{amsmath}
\usepackage{amssymb}

\usepackage{graphicx}
\usepackage{cite}
\usepackage[hyphens]{url}
\usepackage{hyperref}
\usepackage{listings}
\usepackage{siunitx}
\usepackage{tikz}
\usetikzlibrary{arrows}
\usetikzlibrary{patterns}

\IEEEoverridecommandlockouts

\newcommand\copyrighttext{%
\footnotesize \textcopyright \enspace 2019 IEEE. Personal use of this material is permitted. Permission from IEEE must be obtained for all other uses, in any current or future media, including reprinting/republishing this material for advertising or promo
tional purposes, creating new collective works, for resale or redistribution to servers or lists, or reuse of any copyrighted component of this work in other works.  DOI: \href{https://doi.org/10.1109/BlackSeaCom.2019.8812774}{10.1109/BlackSeaCom.2019.8812774}
}
\newcommand\copyrightnotice{%
\begin{tikzpicture}[remember picture,overlay]
\node[anchor=south] at (current page.south) {\fbox{\parbox{\dimexpr\textwidth-\fboxsep-\fboxrule\relax}{\copyrighttext}}};
\end{tikzpicture}%
}

\begin{document}
	
	\title{OFDMA Resource Allocation\\for Real-Time Applications\\in IEEE 802.11ax Networks\thanks{The work was carried out at NRU HSE and supported by the Russian Science Foundation (agreement 18-19-00580)}}
	
	\author{\IEEEauthorblockN{
			Evgeny~Avdotin\IEEEauthorrefmark{1},
			Dmitry~Bankov\IEEEauthorrefmark{1}\IEEEauthorrefmark{2},
			Evgeny~Khorov\IEEEauthorrefmark{1}\IEEEauthorrefmark{2},
			Andrey~Lyakhov\IEEEauthorrefmark{1}\\
		}
		
		\IEEEauthorrefmark{2}National Research University Higher School of Economics, Moscow, Russia \\
		\IEEEauthorblockA{\IEEEauthorrefmark{1}Institute for Information Transmission Problems, Russian Academy of Sciences, Moscow, Russia\\
			Email: avdotin.es@phystech.edu, bankov@iitp.ru, khorov@iitp.ru, lyakhov@iitp.ru\\
		}
	}
	
	\maketitle
	\copyrightnotice
	\begin{abstract}
		Support of real-time applications that impose strict requirements on packet loss ratio and latency is an essential feature of the next generation Wi-Fi networks.
		Initially introduced in the 802.11ax amendment to the Wi-Fi standard, uplink OFDMA  seems to be a promising solution for supported low-latency data transmission from the numerous stations to an access point.
		In this paper, we study how to allocate OFDMA resources in an 802.11ax network and propose an algorithm aimed at providing the delay less than one millisecond and reliability up to 99.999\% as required by numerous real-time applications.
		We design a resource allocation algorithm and with extensive simulation, show that it decreases delays for real-time traffic by orders of magnitude, while the throughput for non-real-time traffic is reduced insignificantly.
	\end{abstract}
	
	\section{Introduction}{\label{sec:intro}}
	Advances in information and telecommunication technologies resulted in emergence and development of Real-Time Applications (RTA).
	They include remote control, industrial automation, online gaming, augmented and virtual reality.
	RTA imposes stringent requirements on telecommunication technologies in terms of reliability (packet loss rate (PLR) up to $10^{-8}$--$10^{-5}$ and latency (packet delivery time up to $1$--\SI{10}{\ms}).
	In wireless networks, such requirements are particularly hard to satisfy\cite{discussion_target_presentation,usecases_presentation, tsn_presentation}.
	In November 2017, the discussion of RTA in Wi-Fi initiated at the IEEE 802 LAN/MAN Standards Committee Plenary Session \cite{tsn_presentation, sage} led to creation of an RTA Topic Interest Group (RTA TIG) with the goal to classify the RTA scenarios essential for Wi-Fi networks, to determine their requirements in terms of latency, PLR and the number of devices served by one Wi-Fi access point (AP), and to propose solutions to enable RTA in Wi-Fi. 
	
	Establishing a corresponding level of reliability and latency in Wi-Fi networks is a complicated problem.
	The reasons for that are related to the default Wi-Fi random access with collisions that result in delays and packet losses.
	Also, Wi-Fi stations (STAs) do not start a transmission when the channel is busy. So high priority urgent frames are to be transmitted only when the ongoing transmission ends.
	
	One of the possible solutions to satisfy RTA requirements for uplink transmission is to use orthogonal frequency division multiple access~(OFDMA) introduced in the Wi-Fi standard, as a part of the IEEE 802.11ax amendment \cite{khorov2018tutorial}.
	OFDMA is essential for RTA since it enables several STAs to transmit or receive data at one time.
	It also allows allocating resource units (RUs) for STAs to transmit without collisions and defines uplink OFDMA-based random access (UORA).
	According to RTA TIG reports, these features seem to be promising for providing high reliability and low latency \cite{performance_presentation}. 
	
	It should be considered that the bandwidth available for non-RTA transmissions is reduced due to the allocation of OFDMA resource units (RUs) for RTA packets.
	So, the solution of RTA should provide the maximal bandwidth for non-RTA traffic while satisfying RTA traffic requirements.
	
	In this paper, we study the aforementioned problems and present a Cyclic Resource Assignment Algorithm (CRA) which can be used to satisfy RTA requirements in IEEE 802.11ax networks.
	We evaluate the efficiency of the proposed algorithm and compare it with the standard 802.11ax UORA approach.
	We show that CRA can meet the RTA requirements, while standard UORA is inappropriate for RTA scenarios.
	
	\section{OFDMA in IEEE 802.11ax}{\label{sec:OFDMA}}
	OFDMA is one of the fundamentals of IEEE 802.11ax.
	It introduces the division of channel resource in the frequency domain, thus extending the standard CSMA/CA.
	Namely, OFDMA allows splitting the channel into several RUs of various size.
	
	Wi-Fi uplink transmissions with OFDMA are established as follows.
	To start a transmission, the AP sends a Trigger Frame (TF), which synchronizes transmitting STAs and informs them about the transmission parameters and RU assignment.
	Then the STAs use the assigned RUs to transmit their data frames.
	Finally, to acknowledge the successfully received packets, the AP transmits a Multi-STA Block Acknowledgment frame (MSBA).
	The TF transmission, RUs and the MSBA are separated by a Short Inter-Frame Space (SIFS).
	
	To transmit a data frame using UORA, a STA uses the subsequent OFDMA Back-off (OBO) procedure.
	The OBO counter of the STA is initialized to a random value equiprobably drawn from the set $\{0, 1, ..., OCW - 1\}$, where $OCW$ is the OFDMA contention window.
	When the STA receives a TF, it compares its OBO counter with the number of RUs allocated for UORA.
	If the OBO counter is greater than the number of RUs, STA decreases its OBO counter by this number and waits for the next TF.
	Otherwise, the STA uses a random RU assigned to UORA.
	
	In case of collision, $OCW$ value of collision STA doubles, unless it reaches the $OCW_{MAX}$ limit.
	If the transmission is successful, the STA sets its $OCW$ value to the minimal value $OCW_{MIN}$.
	AP specifies $OCW_{MIN}$ and $OCW_{MAX}$ parameters in beacons and Probe Response Frames.
	
	\begin{figure}[tb]
		\centering
		\includegraphics[width=\linewidth]{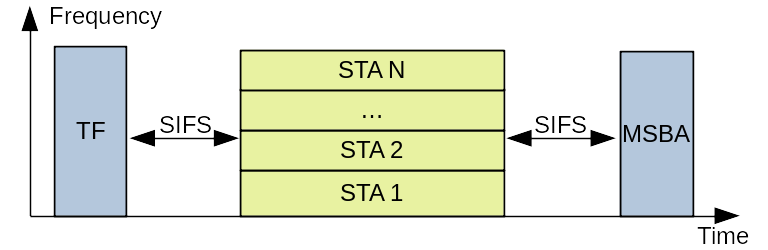}
		\caption{\label{fig:channel} Frame exchange sequence for uplink OFDMA}
	\end{figure}

	\section{Problem Statement}
	\label{sec:problem}
	Consider an IEEE 802.11ax network consisting of an AP, $N$ STAs that transmit RTA frames in uplink and several devices transmitting saturated flows.
	New RTA frames are generated following successful transmissions with the exponentially distributed delay with parameter $\lambda.$
	
	The network uses OFDMA for all transmissions.
	The AP maintains such operation by periodically broadcasting TFs that contain the scheduling information.
	The period of TFs is called a \emph{slot}.
	We consider all the slots having the same duration.
	
	We also assume that the RTA frames are short enough for STA to manage to transmit one frame in one RU.
	
	It is considered that frames can be lost only when two or more STAs transmit in the same RU, i.e., in case of collision.
	In such a case, none of the transmitted frames are delivered.
	
	For the presented scenario, we state the problem to develop an algorithm of resource allocation for RTA frames transmission.
	It shall provide packet transmission delay less than \SI{1}{\ms} with probability 99.999\%.
	At the same time, the portion of consumed channel resources should stay minimal.
	
	\subsection{Cyclic Resource Assignment Algorithm}
	\label{sec:algorithm}
\begin{figure}[tb]
		\centering
		\includegraphics[width=\linewidth]{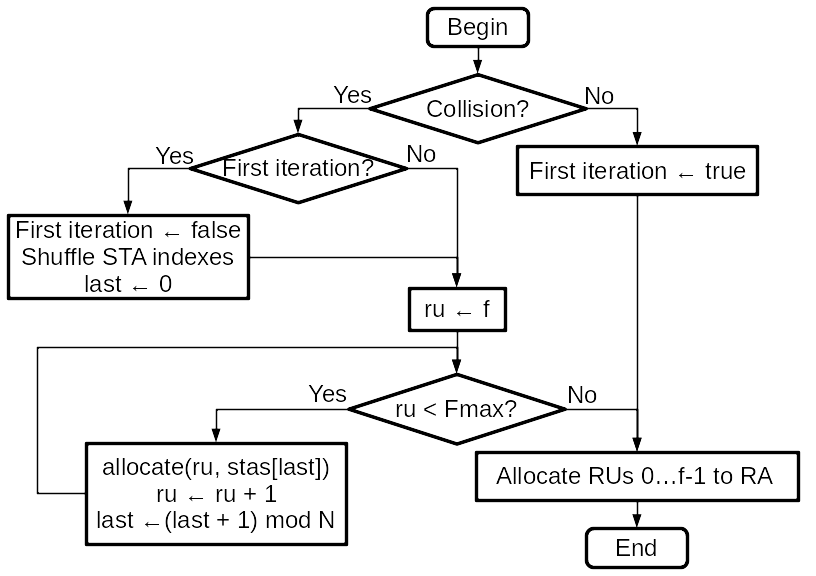}
		\caption{\label{fig:cra}
			Block scheme of CRA.}
	\end{figure}

\begin{figure}[tb]
		\centering
		\includegraphics[width=\linewidth]{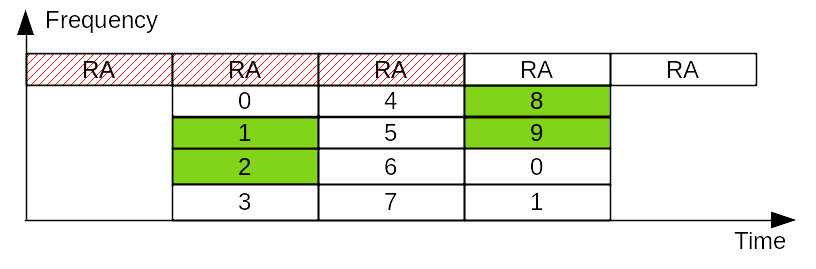}
		\caption{\label{fig:cyclic_assignment}
			An example of CRA operation. STAs 1, 2, 8 and 9 generate frames.}
	\end{figure}

	In this section, we design a resource allocation algorithm called the Cyclic Resource Assignment (CRA), see Fig.~\ref{fig:cra} that presents the block scheme of the algorithm, and Fig.~\ref{fig:cyclic_assignment} which illustrates the operation of the algorithm.
	In Fig.~\ref{fig:cyclic_assignment}, the RUs assigned for RA are marked as ``RA'', numbers are the shuffled indexes of the STAs that are allocated to specific RUs, hatched RUs indicate collisions and green RUs indicate successful transmissions.
	
	The goal of the algorithm is to minimize the RTA data transmission delay.
	
	The algorithm schedules the resources at the beginning of each slot.
	It allocates only the smallest, 26-tone RUs for RTA STAs, thus enabling the AP to allocate the maximal number of RUs and to maximize the number of simultaneously served STAs. Let  $F_{max}$ be the maximal number of RUs that can be allocated in the channel.
	The CRA always allocates $f$ RUs for random access (RA), where $f<F_{max}$ is a parameter of the algorithm.
	Transmissions in RA are performed according to UORA, but $OCW_{MIN}$ and $OCW_{MAX}$ parameters are set to 0. This allows the STA using all the RU allocated for RA is in the nearest slot. 
	
	We consider that the saturated data flows occupy all the available resources which are not allocated for RTA STAs and for RA, so the resource allocation for the saturated data is not mentioned further.
	
	The CRA works as follows. 
	
	If the previous slot has no collisions, in the next slot, the AP does not schedule any RUs for RTA traffic in a deterministic way.
	In such a case, the AP knows that there are no STAs in the network which have urgent data but could not transmit it.
	In the next slot, in $f$ RUs allocated for RA, only those STAs may transmit that have recently generated frames.
	
	In there are some collisions in the previous slot, namely in some RUs allocated for RA, the AP knows that some RTA STAs have frames, but does not know precisely which STAs.
	To determine these STAs and to allow them to resolve the collisions, the AP gives every STA a chance to make a transmission without contention. 
	For this, the AP schedules $F_{max} - f$ RUs for the deterministic access.
	In the first slot after the collision in RA, the AP assigns RUs to the first $F_{max} - f$ STAs, then to the next $F_{max} - f$ STAs and so on.
	As this takes place, STAs are enumerated in a cycle, as shown in Fig.~\ref{fig:cyclic_assignment}.
	If during the cycle there are no collisions in a slot, then the AP stops the cycle and assigns $f$ RUs for RA only.
	
	When a cycle is started, the AP randomly shuffles the STA numbers, thus eliminating a situation when a STA always gets resources before another STA.

	\section{Numerical Results}
	\label{sec:numerical}
	To evaluate the performance of the designed algorithm, we simulate the operation of a Wi-Fi network in the scenario described in Section \ref{sec:problem}.
	In our experiments, the OFDMA slot duration equals \SI{250}{\us}, while the maximum number of the used RUs (the $F_{max}$ parameter) equals 18.
	The RTA STAs transmit frames which are small enough to fit into one RU.
	
\begin{figure}[tb]
		\centering
		\includegraphics[width=\linewidth]{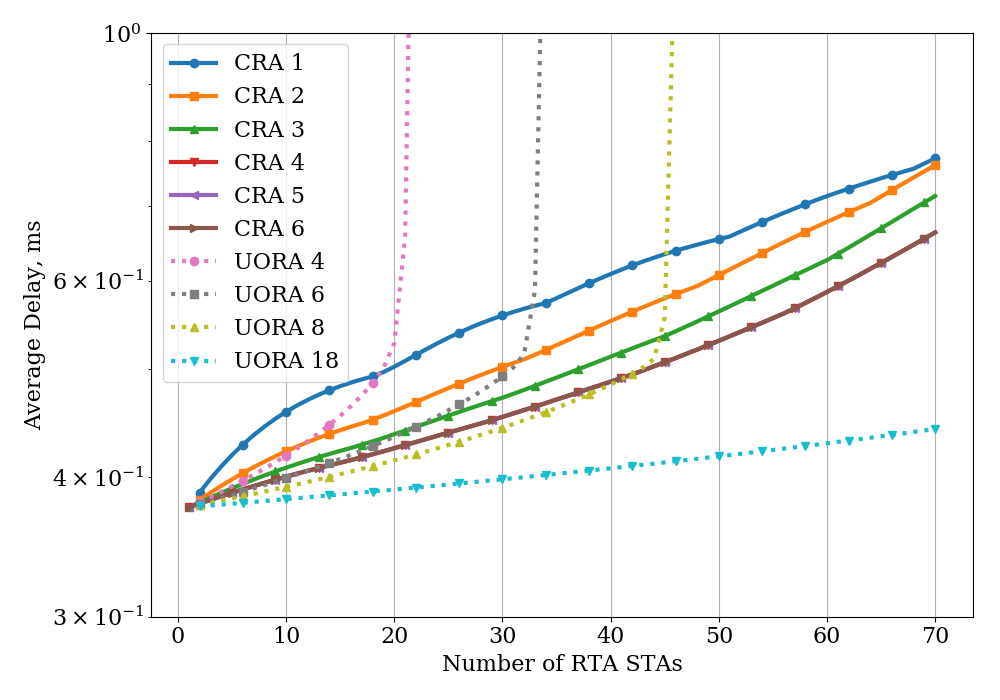}
		\caption{\label{fig:delay}
			The dependency of the average delay for RTA frames on the number of RTA STAs. Packets arrival rate $\lambda = 200 s^{-1}$.}
	\end{figure}

\begin{figure}[tb]
		\centering
		\includegraphics[width=\linewidth]{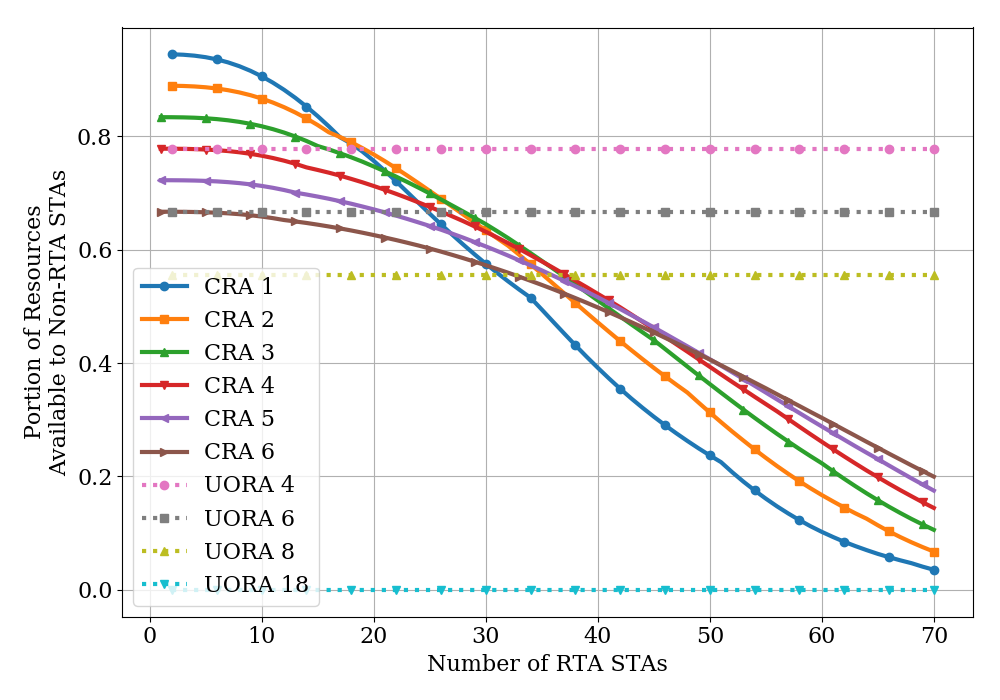}
		\caption{\label{fig:goodput}
			The dependency of the portion of resources available for non-RTA STAs on the number of RTA STAs. Packets arrival rate $\lambda = 200 s^{-1}$.}
	\end{figure}

	We compare CRA with pure UORA. For both algorithms, we vary the number $f$ of RUs assigned to random access. The value of $f$ is shown in the legend in Fig.~\ref{fig:delay} and \ref{fig:goodput}.

	Fig.~\ref{fig:delay} shows the dependency of the average delay for RTA frames on the number of RTA STAs.
	As one can see, this dependency is linear for CRA, but for UORA it is unstable and, when the number of STAs reaches some value, suddenly increases.
	The average delay of RTA frames is lower than \SI{1}{\ms} for a wide range of the numbers of STAs, and is generally lower for larger $f$ values.
	Such a dependency on $f$ is explained by the fact that big $f$ yields a higher probability of successful transmission in RA, and even in case of CRA, the STA can transmit its frame early in RA without the need to wait for its turn to transmit in deterministic access.
	
	\begin{figure}[tb]
		\centering
		\includegraphics[width=\linewidth]{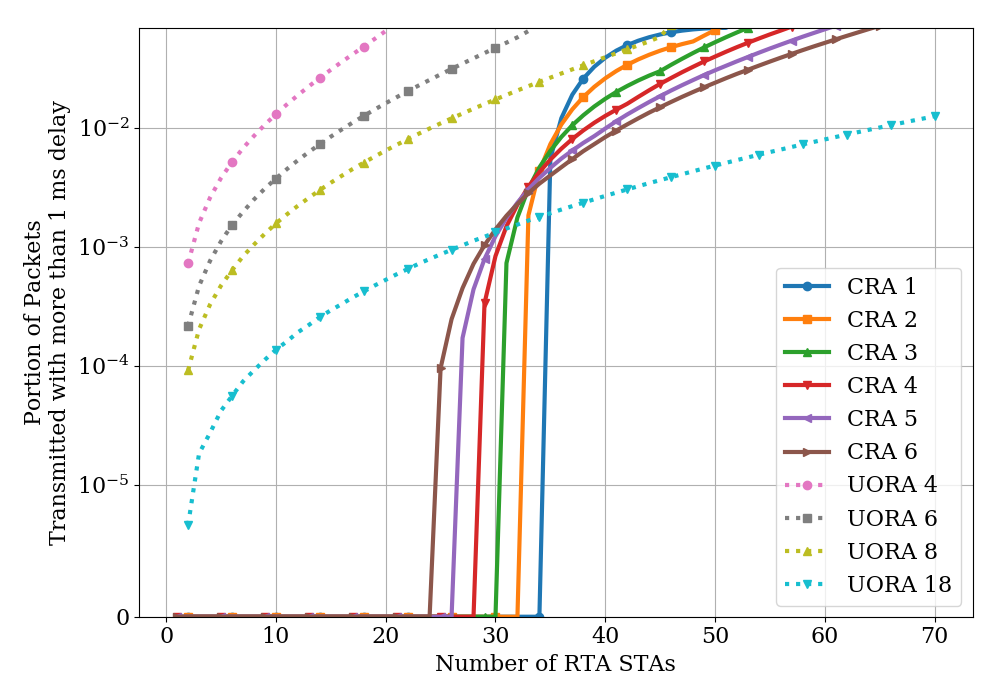}
		\caption{\label{fig:late_ratio}
			Dependency of the portion of packets transmitted with more than \SI{1}{\ms} delay on the number of RTA STAs. Packets arrival rate $\lambda = 200 s^{-1}$.}
	\end{figure}
	
	However, for RTA the average delay is not as significant as the delay percentile.
	Fig.~\ref{fig:late_ratio} shows the dependency of the portion $P_{late}$ of packets delivered with more than \SI{1}{\ms} delay on the number of RTA STAs.
	This plot has a linear scale for $P_{late}$ less than $10^{-5}$ and a logarithmic scale for $P_{late}$ greater than $10^{-5}$.
	The obtained results show that pure UORA performs worse than CRA.
	For UORA, lower $P_{late}$ is achieved with a lower number of RUs used in random access. However, the required probability of $10^{-5}$ is achieved only for a small number of STAs and $f = 18$, i.e., when the whole channel is used for RA.
	For the CRA, the value of $P_{late}$ is zero for the number of STAs lower than $\left(F_{max} - f\right) \cdot 2$ and suddenly increases for greater numbers.
	Such a number of STAs corresponds to the situation when a STA waits too long for deterministic access while collisions in RA do not let it transmit either.
	From the plots, we can see that even though for most algorithms the average delay is less than \SI{1}{\ms}, the CRA is much more efficient in terms of the RTA requirements.
	
	Another performance indicator which should be considered is the proportion of network bandwidth available for non-RTA STAs, the dependency of which on the number of RTA STAs is shown onFig.~\ref{fig:goodput}.
	For pure UORA, it equals $\frac{F_{max} - f}{F_{max}}$ and decreases with the increasing number of RUs assigned to RTA STAs.
	For CRA, the proportion of bandwidth available for non-RTA STAs reduces with the increasing number of STAs, because for a higher number of RTA STAs more time is spent on collision resolution and on the cycling through STAs when all channel resources are used to resolve conflicts.
	
	\section{Conclusion}
	\label{sec:outro}
	In this work, we have studied several methods to allow Real-Time Applications (RTA) in IEEE 802.11ax networks.
	We have shown that with the standard 802.11ax UORA it is almost impossible to satisfy the RTA requirements of \SI{1}{\ms} delay with the probability of 99.999\%.
	We have designed a resource allocation algorithm called CRA which is aimed at satisfying such requirements.
	With extensive simulation, we have shown that this algorithm is much more efficient than UORA and can provide RTA service for a wide range of users.
	Our solution is efficient not only in terms of the delay percentile but also in terms of the amount of channel resource available for usual traffic.
	
	\bibliographystyle{IEEEtran}
	\bibliography{biblio.bib}
	
\end{document}